\documentstyle[epsfig]{elsart}
\begin{document}
\begin{frontmatter}
\title{Period Stabilization in the Busse-Heikes Model of the 
K\"uppers-Lortz Instability\thanksref{joel}}
\author{R. Toral\thanksref{email}}, \author{M. San Miguel and R. Gallego} 
\address{
Instituto Mediterr\'aneo de Estudios Avanzados\thanksref{www} (UIB-CSIC),
E-07071 Palma de Mallorca, Spain
}
\thanks[joel] 
{This paper is dedicated to Joel Lebowitz on the occasion of his 70th birthday.}
\thanks[email]{email: {\tt raul@imedea.uib.es}}
\thanks[www]{Web site: http://www.imedea.uib.es/PhysDept/}

\begin{abstract}
The Busse-Heikes dynamical model is described in terms of relaxational and
nonrelaxational dynamics. Within this dynamical picture a diverging alternating
period is calculated in a reduced dynamics given by a time-dependent Hamiltonian
with decreasing energy. A mean period is calculated which results from noise
stabilization of a mean energy. The consideration of spatial-dependent
amplitudes leads to vertex formation. The competition of front motion around
the vertices and the K\"uppers-Lortz instability in determining an alternating
period is discussed.
\end{abstract}

\end{frontmatter}

\section{Introduction}

One of the most extensively studied systems, in the field of pattern formation
in nonequilibrium systems, is Rayleigh-B\'enard thermal convection. In many
geophysical and astrophysical systems,  thermally induced convection is
combined with Coriolis forces induced by rotation. Therefore, Rayleigh-B\'enard
convection in fluid layers rotating around a vertical axis is a hydrodynamical
system of significant importance. Specially interesting is a spatio-temporal
regime that takes place above a critical rotation angular velocity. The system
breaks up into a persistent dynamical state such that set of parallel
convection rolls are seen to change orientation with a characteristic period.
This phenomenon is known as the K\"uppers-Lortz instability
\cite{KuppersLortz}.  This instability can be described as follows: for  an
angular rotation speed $\Omega$ greater than some critical  value $\Omega_c$,
convective rolls lose their stability with respect to rolls inclined at an
angle of about $60^0$ in the sense of rotation. The new rolls undergo the same
instability, so that there is no stable steady-state pattern. As a result
spatially disordered patterns arise already arbitrarily close to the onset of
convection. Experimental characterization of this regime of spatio-temporal
chaos has been reported in~\cite{KLexp}. Most experiments have been performed
for small Prandtl numbers and have been theoretically described in the realm of
Swift-Hohenberg models with \cite{SHE_mf} and without \cite{SHE_nmf} the
inclusion of mean flow effects. On the other hand, large Prandtl numbers lead to
more rigid convection rolls. In this situation mean flow coupling can be
neglected and, in the limit of infinite Prandtl numbers, three-mode models
have been shown \cite{CrossTu,CrossSHE} to exhibit the same qualitative
features as more sophisticate Swift-Hohenberg models that take into account the
full range of possible roll orientations.

We consider in this paper a three-mode model proposed by Busse and Heikes
\cite{BusseHeikes} to study the K\"uppers-Lortz instability. Each mode
represents the amplitude of a set of parallel rolls with an orientation of
$60^\circ$ to each other. This model contains an attracting heteroclinic cycle
connecting three fixed points corresponding to the three different roll
solutions. The model predicts successfully the existence of a region in
parameter space  in which the roll solution is unstable, but fails to reproduce
the experimental observation of an approximately constant period between roll
alternation. Whereas Busse and Heikes speculated that  such a constant period
would be obtained by the addition of noise, a conclusion confirmed by Stone and
Holmes \cite{stone},  no systematic study of the relation of the period to the
system parameters has been performed so far. Another explanation for period
stabilization has been given by Cross and Tu \cite{CrossTu} who have performed
numerical investigations of an extension of the Busse-Heikes equations, where a
spatial variation of the amplitudes has been introduced. In this paper, we
study in detail these two proposed mechanisms for period stabilization  in the
Busse-Heikes model: (i) addition of noise and (ii) the consideration of
spatial-dependent terms. 

The paper is organized as follows: in section \ref{II} we present a description
of the Busse-Heikes model and give a clear physical explanation of the
period divergence. We describe the dynamics in terms of a relaxational and a
nonrelaxational part. The alternating period is calculated for the latter part
which is associated with a slowly varying time-dependent Hamiltonian. In
section \ref{BHnoise} we consider the same model with the inclusion of additive
white noise terms and we calculate the mean period stabilized by noise in terms
of the previous dynamical picture. Sections \ref{II} and \ref{BHnoise} discuss
ordinary differential equations for the amplitudes of the three modes. In
section \ref{IV} we consider the more physically appropriate situation of
spatial-dependent amplitudes in a $d=2$ model and study the influence on the
dynamics of
isotropic and anisotropic spatial-dependent terms. We describe the formation of
vertices and how the period of roll alternation is determined by the competition
of front motion around the vertices and the K\"uppers-Lortz
instability.

\section{Busse-Heikes Model}
\label{II}

Based on the fact that, in a first approximation, only three directions are
relevant to this problem, Busse and Heikes \cite{BusseHeikes} proposed a
dynamical model to study the K\"uppers-Lortz instability. The vertical
component of the velocity field $\Psi({\mathbf r},t)$ is written as:
\begin{equation}
\Psi({\mathbf r},t) = \sum_{j=1}^3 A_j({\mathbf r},t){\rm e}^{iq_0\hat{\mathbf
e}_j\cdot{\mathbf r}} + {\mathrm c.c.}
\end{equation}
(``c.c." denotes complex conjugate).  The vectors $\hat{\mathbf e}_j$ are unit
vectors in directions $j=1,2,3$ which form an angle of $60^\circ$ between them, and
$q_0$ is the selected wavenumber of the convection pattern. In this model the
(complex) amplitudes of the three rotating modes, $A_1$, $A_2$, $A_3$, are
independent of space and follow the evolution equations \cite{BusseHeikes}:
\begin{eqnarray}
\dot A_1 & = & A_1[\nu-|A_1|^2-(1+\mu+\delta)|A_2|^2-
(1+\mu-\delta)|A_3|^2], \nonumber\\
\dot A_2 & = & A_2[\nu-|A_2|^2-(1+\mu+\delta)|A_3|^2-
(1+\mu-\delta)|A_1|^2], \label{eq:bh}\\
\dot A_3 & = & A_3[\nu-|A_3|^2-(1+\mu+\delta)|A_1|^2-
(1+\mu-\delta)|A_2|^2]. \nonumber
\end{eqnarray}
The parameter $\nu$ is proportional to the difference between the Rayleigh number
and the the critical Rayleigh number for convection. We will consider
exclusively in this paper the case of well-developed convection for which the
parameter $\nu$ can be rescaled to $1$, i.e. $\nu=1$ henceforth. The exact relation
of $\mu$ and $\delta$ to the fluid properties has been given in
\cite{KuppersLortz}. We mention here that  $\mu$ is a parameter related to the
temperature gradient and the Taylor number (proportional to the rotation speed
$\Omega$) in such a way that it takes a nonzero value in the case of no
rotation, $\Omega=0$, whereas  $\delta$ is related to the Taylor number in such
a way  that $\Omega=0$ implies $\delta=0$. We will consider only $\Omega > 0$,
or $\delta > 0$; the case $\Omega < 0$ ($\delta < 0$) follows by a simple
change of the coordinate system. Although the dynamical equations are defined
for all values of the parameters, only the case $\mu \ge 0$ is physically
relevant. Writing $A_j=\sqrt{R_j}{\rm e}^{i\theta_j}$ we obtain equations for
the modulus square of the amplitudes $R_j$:
\begin{eqnarray}
\dot R_1 & = & 2R_1[1-R_1-(1+\mu+\delta)R_2-
(1+\mu-\delta)R_3], \nonumber \\
\dot R_2 & = & 2R_2[1-R_2-(1+\mu+\delta)R_3-
(1+\mu-\delta)R_1], \label{bhr}\\
\dot R_3 & = & 2R_3[1-R_3-(1+\mu+\delta)R_1-
(1+\mu-\delta)R_2], \nonumber 
\end{eqnarray}
and for the phases $\theta_j$:
\begin{eqnarray}
\dot \theta_1 & = & 0, \nonumber\\
\dot \theta_2 & = & 0, \label{bha}\\
\dot \theta_3 & = & 0. \nonumber
\end{eqnarray}
It follows that the phases are simply arbitrary constants fixing the location
of the rolls. A solution of the  form $\Psi({\mathbf r})=\sqrt{R_j}{\rm
e}^{i(q_0\hat{\mathbf e}_j\cdot{\mathbf r}+\theta_j)}+{\mathrm c.c.}$ 
represents a set of rolls of wavelength $2\pi/q_0$, oriented in a direction
perpendicular to the vector $\hat{\mathbf e}_j$. Hence, in this model one can
simply consider the equations for the real variables $R_j$ instead of the
equations for the complex variables $A_j$.  A similar set of equations has been
proposed to study population  competition dynamics. For a single biological
species,  the Verhulst or logistic model assumes that  its population $N(t)$
satisfies the evolution equation:
\begin{equation}
\frac{dN}{dt} = r N (1-\lambda N),
\end{equation}
where $r$ is the reproductive growth rate and $\lambda$ is a coefficient
denoting competition amongst the members of the species. If three species are
competing together, it is adequate in some occasions to model this competition
by introducing a Gause--Lotka--Volterra type of equations \cite{ML,bennaim}:
\begin{eqnarray}
\dot N_1 & = & r N_1\left(1-\lambda N_1-\alpha N_2-\beta N_3\right), \nonumber \\
\dot N_2 & = & r N_2\left(1-\lambda N_2-\alpha N_3-\beta N_1\right), \\
\dot N_3 & = & r N_3\left(1-\lambda N_3-\alpha N_1-\beta N_2\right), \nonumber 
\end{eqnarray}
which are the same that the Busse-Heikes equations (\ref{bhr}) for the modulus
square of the amplitudes $R_j$ with the identifications: $r=2$, $\lambda=1$,
$\alpha=1+\mu+\delta$, $\beta=1+\mu-\delta$. These  equations are the basis of
May and Leonard analysis \cite{ML}. We also mention the work of
Soward \cite{Soward} which is concerned with the study of the nature of the
bifurcations mainly, but not limited to, close to the convective instability
for small $\nu$, in a slightly more general model that includes also quadratic
nonlinearities in the equations. In the remaining of the section we will
analyze some of the properties of the solutions of the Busse-Heikes equations
(\ref{eq:bh}). Although our analysis essentially reobtains the results of May
and Leonard, we find it convenient to give it in some detail because, besides
obtaining some further analytical expressions for the time variation of the
amplitudes, we are able in some cases of rewriting the dynamics in terms of a
Lyapunov potential. The existence of this Lyapunov potential allows us to
interpret the asymptotic dynamics for $\mu=0$ as a residual (conservative)
Hamiltonian dynamics. For $\mu >0$ we will use an adiabatic approximation with
a time-dependent Hamiltonian. This interpretation will turn out to be very
useful in the case that noise terms are added to the dynamical equations,
because the found Lyapunov potential governs approximately the stationary
probability distribution. 

We first look for stationary solutions of the Busse-Heikes equations
(\ref{eq:bh}). The fixed point solutions are the following:

(a) The {\sl null} solution: $R_1=R_2=R_3=0$.

(b) {\sl Roll} solutions. There are three families of these solutions, each
characterized by a unique nonvanishing amplitude, for instance:
$(R_1,R_2,R_3)=(1,0,0)$ is a roll solution with rolls perpendicular to the
$\hat{\mathbf e}_1$ direction,  and so on.

(c) {\sl Hexagon} solutions. The three amplitudes are equal and different from 0, namely 
$R_1=R_2=R_3= \frac{1}{3+2\mu}$. They only exist for $\mu > -3/2$.

(d) {\sl Rhombus} solutions. There are three families of these solutions, in which 
two amplitudes are different from $0$ and the
third amplitude vanishes. For instance:
$(R_1,R_2,R_3)=(\frac{\mu+\delta}{\mu(\mu+2)-\delta^2},
\frac{\mu-\delta}{\mu(\mu+2)-\delta^2},0)$.
They only exist for $\mu > \delta$, or  $-1-\sqrt{1+\delta^2} < \mu < 
- \delta$.

\begin{figure}
\begin{center}
\psfig{figure=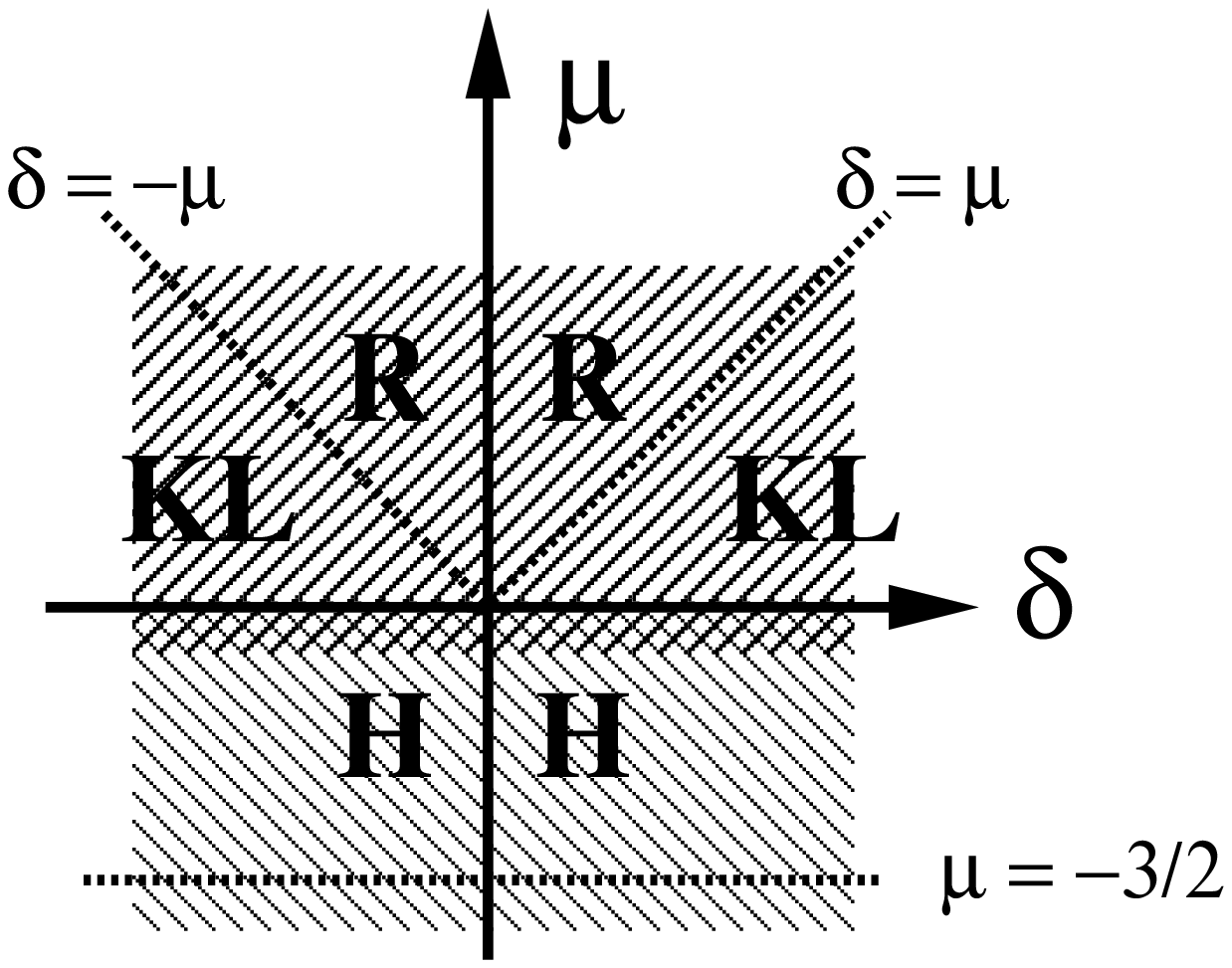,width=0.6\textwidth}
\end{center}
\caption{\label{fig1}
Stability regions for the Busse-Heikes dynamical system 
(\ref{eq:bh}). The region `H' is where the hexagon solution (three equal
amplitudes) is stable. In the `R' region, the three roll
solutions are stable, and in region `KL' there are 
no stable fixed points.}
\end{figure}

The stability of the previous solutions can be studied by  means
of  a linear stability analysis.  The result is summarized in  Fig.
\ref{fig1}.  For $\mu < -3/2$ there are no stable solutions and the amplitudes
grow without limit. The rhombus and null solutions are never stable.  The
hexagon solutions are stable for $-3/2 < \mu < 0$. The roll solutions are stable
for $\mu > \delta$. For $0 < \mu < \delta$ there are no stable solutions, but 
the amplitudes remain bounded. This instability can be described as follows:
consider the unstable roll solution $(R_1,R_2,R_3) = (1,0,0)$. The amplitude of the $A_2$ mode starts growing and that of $A_1$
decreasing in order to reach the roll solution $(0,1,0)$. However, this new
roll solution is also unstable, and before it can be reached, the dynamical
system starts evolving towards the roll solution $(0,0,1)$, which is unstable
and evolves towards the solution $(1,0,0)$ which is unstable, and so on.
Schematically, we can represent the situation as follows:
\begin{equation}
(1,0,0) \to (0,1,0) \to (0,0,1) \to (1,0,0) \to (0,1,0) \dots
\end{equation}
This is the K\"uppers-Lortz instability that 
shows up in the rotation of the convective rolls.
The K\"uppers-Lortz unstable region is characterized by the presence
of three unstable fixed points, and a
heteroclinic cycle connecting them.

The novelty of our treatment
consists in writing the Busse-Heikes equations of motion in the form:
\begin{equation}
\label{dotap}
\dot A_j = -\frac{\partial V}{\partial A_j^\ast} + \delta \, v_j, ~~~~~~j=1,2,3,
\end{equation}
with the {\sl potential function}:
\begin{eqnarray} 
V(A_1,A_2,A_3) & = & -\left(|A_1|^2+|A_2|^2+|A_3|^2\right) + \frac{1}{2}
\left(|A_1|^4+|A_2|^4+|A_3|^4\right)+\nonumber\\
& & (1+\mu)\left(|A_1|^2|A_2|^2+
|A_2|^2|A_3|^2+|A_3|^2|A_1|^2\right) \label{potential} \\
& = & -\left(R_1+R_2+R_3\right) + \frac{1}{2}
\left(R_1^2+R_2^2+R_3^2\right)+(1+\mu)\left(R_1R_2+
R_2R_3+R_3R_1\right), \nonumber
\end{eqnarray}
and 
\begin{eqnarray}
\label{vi}
v_1& =& A_1(-|A_2|^2+|A_3|^2) = A_1(-R_2+R_3), \nonumber \\
v_2& =& A_2(-|A_3|^2+|A_1|^2) = A_2(-R_3+R_1), \\
v_3& =& A_3(-|A_1|^2+|A_2|^2) = A_3(-R_1+R_2). \nonumber
\end{eqnarray}
The first term in the right-hand side of (\ref{dotap}) describes relaxation in
the potential $V(A_1,A_2,A_3)$. In the case $\delta=0$, hence, the dynamics is
described simply as the relaxation, along the gradient lines of the potential
$V$, in order to reach a minimum of $V$. In the case $\delta >0$ there is
another contribution to the dynamics. Its effect can be analyzed partly by
looking at the time evolution of the potential:
\begin{equation}\label{dvdta}
\frac{dV}{dt}=\sum_{j=1}^3\frac{\partial V}{\partial
A_j}\frac{dA_j}{dt}+{\mathrm c.c.} = -2\sum_{j=1}^3\left|\frac{\partial
V}{\partial A_j}\right|^2 + \delta \left[\sum_{j=1}^3\frac{\partial V}{\partial
A_j}v_j+{\mathrm c.c.}\right].
\end{equation}

\noindent Therefore, when the so-called {\sl orthogonality condition} is
satisfied:
\begin{equation}\label{ortho}
  \delta\left[\sum_{j=1}^3\frac{\partial V}{\partial A_j}v_j+{\mathrm c.c.}\right]=0,
\end{equation}

\noindent the function $V$ decreases along the dynamical trajectories and it
becomes a Lyapunov potential \cite{lyapunov} if it is bounded from below (which
is the case for $\mu>-3/2$). Using equations (\ref{potential})
and (\ref{vi}), (\ref{dvdta}) can be rewritten as:
\begin{equation}
\label{dVdt}
\frac{dV}{dt}=-2\sum_{j=1}^3\left|\frac{\partial V}{\partial A_j}\right|^2 -2
\mu \delta (|A_1|^2-|A_2|^2)(|A_2|^2-|A_3|^2)(|A_3|^2-|A_1|^2),
\end{equation}

\noindent so that the orthogonality condition is seen to be satisfied for
$\mu\delta=0$. In the case $\delta=0$, the system is purely relaxational in the
potential $V$ and the corresponding stability diagram can be obtained also by
looking at the minima of $V$. For the null solution, the potential takes the
value $V=0$; for the rhombus, $V=-1/(2+ \mu)$; for the roll solution, $V=
-1/2$; and, finally, for the hexagon solution,  $V=-3/(6+4\mu)$. The study of
the potential (for $\delta=0$) shows that the null and rhombus solutions
correspond always to maxima of the potential and are, therefore, unstable
everywhere. It turns out that the rolls (hexagons) are maxima (minima) of the
potential for $\mu <0$ and minima (maxima) for $\mu>0$. Also, the potential for
the roll solution is smaller that the potential for the other solutions
whenever $\mu>0$, indicating that the rolls are the most stable (and indeed the
only stable ones) solutions in this case. Unfortunately, this  simple criterion
does not have an equivalent in the nonrelaxational case,  $\delta> 0$, for
which one has to perform the full linear stability analysis.

\subsection{The case $\mu=0$}

According to the result (\ref{dVdt}), the function $V(A_1,A_2,A_3)$ is a
Lyapunov potential whenever $\mu \delta=0$. As discussed in the previous
section, the case $\delta=0$ implies a relaxational gradient dynamics in which
all variables tend to fixed values.  In the case $\mu=0$, $\delta >0$, the
dynamics is nonrelaxational potential \cite{msmmahg,mhgmsm,SMT} and, whereas
the dynamics still leads to the surface of minima of the Lyapunov function,
there is a residual motion in this surface for which $dV/dt=0$. In other words:
the relaxational terms in the dynamics make the system evolve towards the
degenerate minimum of the potential (which for $\mu=0$ occurs at
$R_1+R_2+R_3=1$).  The residual motion is governed by the nonrelaxational part
which is proportional to $\delta$ and this residual motion disappears for
$\delta=0$, the relaxational gradient case.

According to this reduction of the dynamics as a residual motion in the surface
of minima of the potential $V$, strictly valid only for $\mu=0$, it turns out
that it is possible to solve essentially the equations of motion. By
``essentially" we mean that after a transient time in which the system is
driven to the minima of $V$, the residual motion is a conservative one in which
it is possible to define a Hamiltonian-like function that allows one to find
explicit expressions for the time variation of the dynamical variables.  Let us
define the variable 
\begin{equation}
X(t)=R_1+R_2+R_3.
\end{equation}
It is straightforward to show
that, for arbitrary $\mu$ and $\delta$, $X$ satisfies the evolution equation:
\begin{equation}
\dot X = 2X(1-X)-4\mu Y,
\end{equation}
where:
\begin{equation}
Y(t)=R_1R_2+R_2R_3+R_3R_1.
\end{equation}
In the case $\mu=0$ the equation for $X(t)$ is a closed equation 
whose solution is 
\begin{equation}
X(t) = \frac{1}{\left( \frac{1}{X_0}-1 \right) {\rm e}^{-2t} +1}.
\end{equation}
Here $X_0=X(t=0)$. From this expression it turns out that  $\lim_{t \to \infty}
X(t) = 1$ independently of the initial condition. In practice, and due to the
exponential decay towards $1$ of the above expression, after a transient time
of order $1$, $X(t)$ already takes its asymptotic value $X(t)=1$. Therefore, we
can substitute $R_1(t)$, say, by $1-R_2(t)-R_3(t)$ to obtain evolution
equations for $R_2(t)$ and  $R_3(t)$. In this way, the original 3-variable
problem, Eqs. (\ref{bhr}), is reduced to a residual dynamics in a 2-variable subspace:
\begin{eqnarray}
\dot R_2 & = &  2\delta R_2(1-R_2-2R_3), \\
\dot R_3 & = & - 2\delta R_3(1-2R_2-R_3). 
\end{eqnarray}
These are Hamilton's equations:
\begin{eqnarray}
\dot R_2 & = & 2\delta \frac{\partial {\cal H}}{\partial R_3}, \\
\dot R_3 & = & -2\delta \frac{\partial {\cal H}}{\partial R_2},
\end{eqnarray}
corresponding to the Hamiltonian:
\begin{equation} 
\label{hamiltonian}
{\cal H}(R_2,R_3) = R_2R_3(1-R_2-R_3).
\end{equation}
As a consequence, in the asymptotic dynamics for which the Hamiltonian
description is valid, ${\cal H}(t)$ is a constant of motion, ${\cal H}=E$,
which will be called the ``energy". The Hamiltonian dynamics is valid
only after a transient time, but the value of $E$ depends only on initial
conditions at $t=0$. The dependence of $E$ on the initial conditions can be
found by introducing the variable $\hat {\cal H}$:
\begin{equation}
{\hat {\cal H}} = R_1 R_2 R_3
\end{equation}
which, in the asymptotic limit ($t\to\infty$) is equivalent to $\cal H$. 
It is easy to show that, for arbitrary values of $\mu$ and $\delta$,
$\hat {\cal H}$ satisfies the following evolution equation:
\begin{equation}
{\hat {\cal H}}^{-1}\frac{d \hat {\cal H}}{dt} = 6-(6+4\mu)X
\end{equation}
(one can reduce the original dynamical problem to variables
$\{X,~Y,~\hat{\cal H}\}$  but the equation for $\dot Y$ turns out to be too
complicated, see \cite{Soward}). If we substitute the solution for $X(t)$ valid
in the case $\mu=0$ we obtain:
\begin{equation}
\hat{\cal H}(t)= \hat {\cal H}_0\left[(1-X_0)e^{-2t}+X_0\right]^{-3},
\end{equation}
with $\hat {\cal H}_0=\hat {\cal H}(t=0)$.
The asymptotic value for ${\cal H}$ is 
\begin{equation}
E = \lim_{t \to \infty} {\cal H}(t) = \lim_{t \to \infty} \hat{\cal H}(t)= 
\frac{\hat{\cal H}_0}{X_0^3} = 
\frac{R_1(0)R_2(0)R_3(0)}{(R_1(0)+R_2(0)+R_3(0))^{3}}.
\end{equation}
Again, this asymptotic value is reached after a transient time of 
order 1. This expression suggests to define the time-dependent variable:
\begin{equation} 
\label{defet}
E(t) = \frac{\hat {\cal H}}{X^3}=\frac{ R_1 R_2 R_3}{(R_1+R_2+R_3)^3},
\end{equation}
whose evolution equation (again, for arbitrary $\mu$, $\delta$) is:
\begin{equation}
\label{dqdt}
\frac{dE}{dt} = -4\mu \left(X-3\frac{Y}{X}\right) E \equiv -4\mu f(t) E.
\end{equation}
Therefore, in the case $\mu=0$, $E(t)=E$ is a constant of motion that
coincides, in the asymptotic limit when $X=1$, with the numerical value of the
Hamiltonian $\cal H$. According to their definition, $E(t)$ is a bounded
function $0 \le E(t) \le 1/27$ and $f(t)\ge 0$ for $R_j\geq 0,~j=1,2,3$.

\begin{figure}[h]
\begin{center}
\psfig{figure=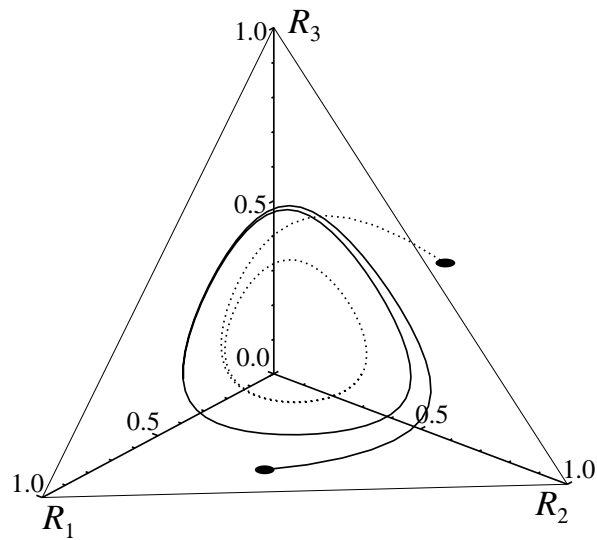,width=0.6\textwidth}
\end{center}
\caption{\label{fig2} Dynamics for $\mu=0$ in the variables $R_1,~R_2,~R_3$ for
two different initial conditions.
After a transient time of order $1$ the motion is on the plane 
$R_1+R_2+R_3=1$}
\end{figure}

\begin{figure}[h]
\begin{center}
\psfig{figure=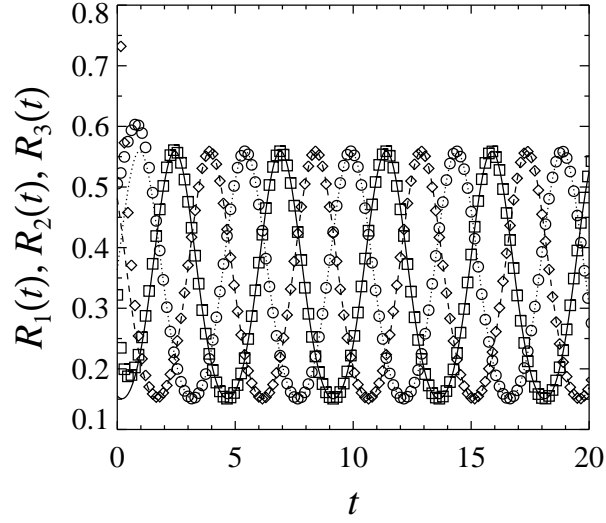,width=0.6\textwidth}
\end{center}
\caption{\label{fig3} 
Time evolution of amplitudes in the case $\delta =1.3$, $\mu=0$. After a
transient time of order $1$, the three variables $R_1,~R_2,~R_3$ vary
periodically in time. The lines are the theoretical predictions that come from
Eqs. (\ref{r3t},\ref{r1t},\ref{r2t}).}
\end{figure}

The problem in the case $\mu=0$ can now be given an explicit solution. 
After a transient time (or order 1), the motion occurs on
the plane $R_1+R_2+R_3=1$, see Fig. \ref{fig2}. The motion
is periodic because it corresponds to 
a Hamiltonian orbit with a fixed energy. The exact shape of the trajectory 
depends on the value of
the energy $E$ which, in turn, depends on initial conditions. More
interestingly, the period of the orbit can also be computed. For this,
we solve the evolution equation (again asymptotically) for, say, $R_3$.
By elimination of $R_2$ by setting ${\cal H}=E$ in Eq. (\ref{hamiltonian}):
\begin{equation} 
R_2= \frac{1}{2}\left( 1-R_3 \pm \sqrt{(1-R_3)^2-4E/R_3}\right),
\end{equation}
we obtain a closed equation for $R_3$:
\begin{equation}
\label{dotr3}
\dot R_3 = \pm 2\delta \sqrt{R_3^2(1-R_3)^2-4ER_3}.
\end{equation}
Let $b$ and $c$ be the return points, i.e. the solutions of 
\begin{equation}
R_3(1-R_3)^2-4E =0,
\end{equation} 
lying in the interval $(0,1)$ \cite{r3ne0}. The three roots, $a,b,c$, of the
above third-degree  equation are real and two of them (the return points
$b,c$)  lie in the interval $(0,1)$. The explicit expression for the roots is:
\begin{eqnarray}
a  & = & \frac{2}{3}\left[1+\cos{\frac{\theta}{3}}\right], \\
b  & = & \frac{2}{3}\left[1+\cos{\frac{\theta-2\pi}{3}}\right], \\
c  & = & \frac{2}{3}\left[1+\cos{\frac{\theta+2\pi}{3}}\right],
\end{eqnarray}
where 
\begin{equation}
\theta=\arccos{(54E-1)}.
\end{equation}
Integration of (\ref{dotr3}) yields the equation of motion for $R_3(t)$:
\begin{equation}
\int_c^{R_3(t)} \frac{dx}{\sqrt{x(x-a)(x-b)(x-c)}} = 2 \delta \int_{t_0}^t dt',
\end{equation}
where we have chosen the initial time $t_0$ to correspond to the minimum value 
when $R_3(t)=c$. The integral in the left hand side can be expressed in terms
of the Jacobi elliptic function \cite{abra} ${\rm sn}[x|q]$, to yield:
\begin{equation}
R_3(t)=\frac{bc}{b+(c-b){\rm sn}^2[\delta\sqrt{b(a-c)} (t-t_0)|q]},
\end{equation}
where
\begin{equation}
q  = \frac{a(b-c)}{b(a-c)}.
\end{equation}
The period of the orbit $T$ can be expressed in terms of the complete elliptic
function of the first kind $K(q)$:
\begin{equation}
\label{te2}
T= \frac{2}{\delta\sqrt{b(a-c)}} K(q)
\end{equation}
and $R_3(t)$ can be written as:
\begin{equation}
\label{r3t}
R_3(t)=\frac{bc}{b+(c-b){\rm sn}^2\left[
\frac{2K(q)}{T}(t-t_0)|q\right]},
\end{equation}
Finally, the evolution equations for the other variables are:
\begin{eqnarray}
R_1(t)& = & R_3(t-T/3), \label{r1t}\\
R_2(t)& = & R_3(t-2T/3). \label{r2t}
\end{eqnarray}
Summarizing, the behavior of the dynamical system in the case $\mu=0$  can be
described as follows: after a transient time (or order $1$) the three variables
$R_1$, $R_2$, $R_3$ vary periodically in time on the plane $R_1+R_2+R_3=1$.
When $R_1$ decreases, $R_2$ increases, etc. The period of the orbit depends
only on the initial conditions through a constant of motion $E$.  The explicit
expression for the period, Eq. (\ref{te2}), shows that  the period
diverges logarithmically when $E$ tends to zero, namely
\begin{equation}
\label{te}
T(E) = -\frac{3}{2\delta}\ln E \times (1+O(E)),
\end{equation} 
and the amplitude of the oscillations $\Delta\equiv b-c$ depends
also on the constant $E$. When $E$ tends to $0$ the amplitude approaches
$1$:
\begin{equation}
\label{bce}
\Delta = (1-2E^{1/2}) \times (1+O(E)).
\end{equation}
All these relations have been confirmed by a numerical integration of the
Busse-Heikes equations. In Fig. \ref{fig3} we plot the time evolution of
the amplitudes in the case $\mu=0$, $\delta=1.3$. In this figure we can observe
that, after an initial transient time, there is a periodic motion
(characteristic of the K\"uppers-Lortz instability) well described by the
previous analytical expressions. 

\subsection{The case $\mu > 0$}
 
Once we have understood the case $\mu=0$, we now turn to $\mu > 0$. In this
case, the function $V$ is no longer a Lyapunov potential and we can not
reduce the motion to a Hamiltonian one on the surface of minima of $V$. However, since the main
features of the K\"uppers-Lortz dynamics are already present in the case
$\mu=0$ one would like to perform some kind of perturbative analysis valid for
small $\mu$ in order to characterize the K\"uppers-Lortz instability. We
exploit these ideas in order to develop some heuristic arguments that will
allow us to make some quantitative predictions about the evolution of the
system.

According to Eq. (\ref{dqdt}), one can infer that $E(t)$ decreases with time in
a characteristic time scale of order $\mu^{-1}$. If $\mu$ is small, $E(t)$
decreases very slowly and we can extend the picture of the previous section by
using an adiabatic approximation. We assume, then, that the evolution for
$\mu>0$ can be described by a Hamiltonian dynamics with an energy that slowly
decreases with time. Hence, in reducing the energy, the system evolves by
spiraling from a periodic orbit to another (similarly to a damped harmonic
oscillator). Assuming this picture of a time-dependent energy $E(t)$, the main
features of the case $\mu=0$ can now be extended. This model has several
predictions:\\

\begin{figure}[h]
\begin{center}
\psfig{figure=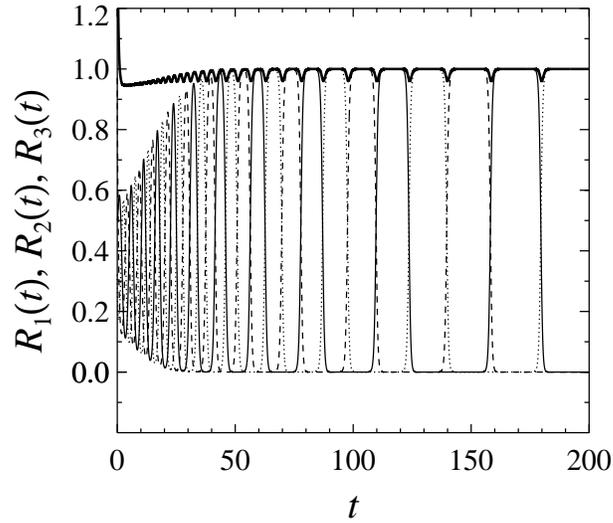,width=0.6\textwidth}
\end{center}
\caption{\label{fig4}  Time evolution of amplitudes in the case $\delta
=1.3$, $\mu=0.1$. The characteristic alternation time of the three variables
$R_1,~R_2,~R_3$ increases with time. Notice that the envelope of the amplitudes
approaches one asymptotically, and that their sum, $R_1+R_2+R_3$ is
approximately equal to $1$.}
\end{figure}

$\bullet$ After a transient time of order $1$, the motion occurs near the plane
$R_1+R_2+R_3 =1$. This is checked in the simulations as we can see in Fig.
\ref{fig4} where we plot the time evolution of the three amplitudes as well as
their sum, in the case $\delta=1.3,~\mu=0.1$.\\

\begin{figure}[h]
\begin{center}
\psfig{figure=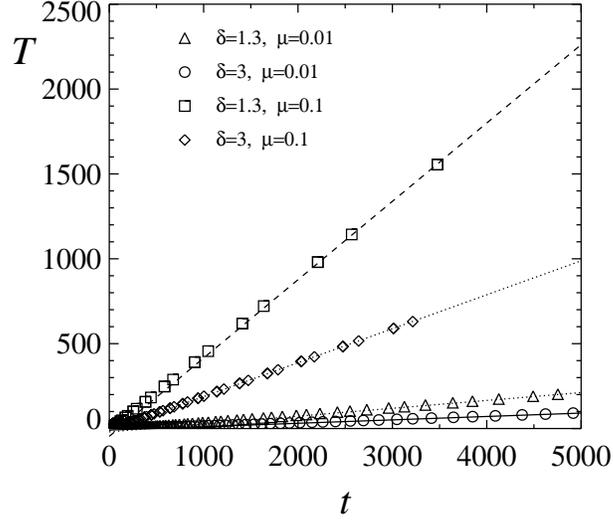,width=0.6\textwidth}
\end{center}
\caption{\label{fig5} 
Time evolution of the period, defined as the time it takes a given amplitude
to cross the reference level $R_j=0.5$ plotted versus time for several values
of  $\delta$ and $\mu$. We also plot straight lines with slopes $\frac{6
\mu}{\delta}$ as predicted by Eq. (\ref{tmd}).}
\end{figure}

$\bullet$ The period of the orbits is now a function of time. Since the energy
decreases towards zero, it follows from Eq. (\ref{te}) that the period
diverges with time. Moreover, it is possible to give an approximate expression
for the time dependence of the period. By integration of equation (\ref{dqdt}),
we obtain:
\begin{equation}
\label{et}
E(t) = E(t_0){\rm e}^{-4\mu\int_{t_0}^{t}f(t')dt'} \approx E(t_0){\rm e}^{-4
\mu (t-t_0)},
\end{equation} 
where we have approximated $f(t)$ by its asymptotic value $f(t)=1$. Once we
have the time evolution of the energy, we can compute the time dependence of
the period by using $T(t)=T(E(t))$ as given by (\ref{te2}). For late times, the
energy is small and the asymptotic result (\ref{te}) leads to:
\begin{equation}
\label{tmd}
T(t) = T_0+\frac{6 \mu}{\delta} t.
\end{equation}
This shows that the period increases linearly with time, in agreement with the
results of \cite{ML} in which the residence period was shown to behave also
linearly with time (although with a different prefactor). In order to check
this relation, we have performed a numerical integration of Eqs. (\ref{eq:bh})
and computed the period $T$, defined as the time it takes for a given amplitude
to cross a reference level (taken arbitrarily as $R_j=0.5$), as a function of
time. The results for $\delta=\{1.3,~3\}$ and $\mu=\{0.1,~0.01\}$, plotted in
Fig. \ref{fig5}, show that there is a perfect agreement between the
theoretical expression and the numerical results.\\

$\bullet$ The amplitude of the oscillations, as given by the return points
$\Delta(t)=b(t)-c(t)$ is now a function of time. Using expression (\ref{bce})
with an energy that decreases with time as in Eq. (\ref{et}) we obtain that the
amplitude of the oscillations increases  with time, see Fig. \ref{fig4},
and that it approaches $1$ in a time or order $t \sim \mu^{-1}$. More
specifically, we have:
\begin{equation}
1-\Delta(t)= (1-\Delta_0){\rm e}^{-2 \mu t}.
\end{equation}

In summary, for the case $\mu>0$, the period of the orbits,  which is a
function of the energy, increases linearly with time and the amplitude of the
oscillations approaches $1$.  We characterize in this way the increase of the
period between successive alternation of the dominating modes, see Fig.
\ref{fig4}, as an effect of the Hamiltonian dynamics with a slowly decreasing
energy. This prediction of the Busse--Heikes model for the K\"uppers-Lortz
instability is unphysical, since the experimental results do not show such an
increase of the period. Busse and Heikes  were fully aware of this problem and
suggested that noise terms (``small amplitude disturbances"), that are present
at all times, prevent the amplitudes from decaying to arbitrary small levels and
a motion which is essentially periodic but with a fluctuating period is 
established. In the next section we study the effect of noise in the dynamical
equations.

\section{Busse-Heikes model in the presence of noise}
\label{BHnoise}

\begin{figure}[h]
\begin{center}
\psfig{figure=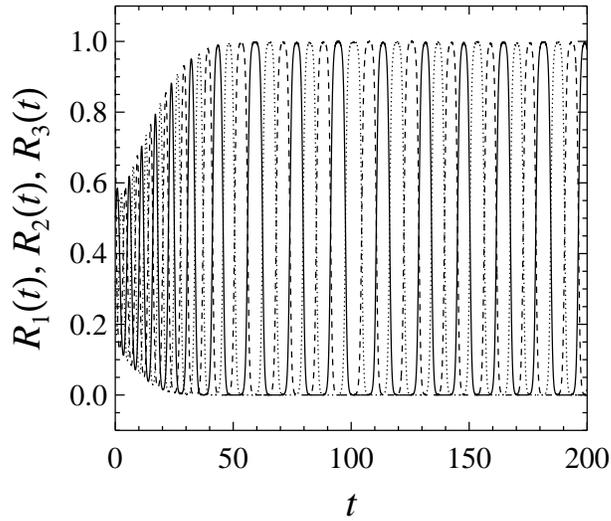,width=0.6\textwidth}
\end{center}
\caption{\label{fig6} 
Time evolution of amplitudes in the presence of noise for $\delta =1.30$,
$\mu=0.1$,  $\epsilon = 10^{-7}$. In this case, the motion is such that the
time interval between dominations of a single mode fluctuates around a mean
value (compare with the equivalent deterministic case shown in Fig.
\ref{fig4}).}
\end{figure}

In order to account for the effect of the fluctuations, we modify the
Busse-Heikes equations by the inclusion of noise terms:
\begin{eqnarray}
\dot A_1 & = & A_1[1-|A_1|^2-(1+\mu+\delta)|A_2|^2-
(1+\mu-\delta)|A_3|^2] + \xi_1(t), \nonumber\\
\dot A_2 & = & A_2[1-|A_2|^2-(1+\mu+\delta)|A_3|^2-
(1+\mu-\delta)|A_1|^2] + \xi_2(t), \label{eq:bhn}\\
\dot A_3 & = & A_3[1-|A_3|^2-(1+\mu+\delta)|A_1|^2-
(1+\mu-\delta)|A_2|^2] + \xi_3(t). \nonumber
\end{eqnarray}
We take the simplest case in which the $\xi_i(t)$ are, complex, white--noise
processes \cite{vanKampen} with correlations:
\begin{equation}
\langle \xi_i(t) \xi_j^\ast(t') \rangle = 2 \epsilon \delta(t-t') \delta_{ij}.
\end{equation}
As mentioned before, and in the case of parameter values lying inside the
K\"uppers-Lortz instability region, noise prevents the system from spending an
increasing amount of time near any of the (unstable) fixed points.  The
mechanism for this is that fluctuations are amplified when the trajectory comes
close to one of the (unstable) fixed points of the dynamics and the trajectory
is then repelled towards another fixed point \cite{stone}. Hence, a
fluctuating, but periodic on average, trajectory is sustained by noise.
Within the general picture developed in the previous section, the main role of
noise for $\mu>0$ is that of preventing $E(t)$ from decaying to zero. This can
be understood in the following qualitative terms: when noise is absent, the
dynamics brings the system to the surface of minima of $V$, where the
dissipative terms act by decreasing the energy in a time scale of order
$\mu^{-1}$, see Eq. (\ref{et}). The inclusion of noise has the effect of
counteracting this energy decrease that occurs in the surface of minima of $V$. As a consequence, $E(t)$ no longer
decays to zero but it stabilizes  around a mean value $\langle E \rangle$. By
stabilizing the orbit around that one corresponding to the mean value $\langle
E \rangle$, fluctuations in the residual motion stabilize the mean period to a
finite value.  In order to check this picture, we have performed numerical
simulations of Eqs. (\ref{eq:bhn}) for small noise amplitude $\epsilon$, using
a stochastic Runge-Kutta algorithm \cite{SMT}. The numerical simulations, see
Fig. \ref{fig6}, show indeed that the trajectories have a well defined
average period $\langle T \rangle$. 

\begin{figure}
\begin{center}
\psfig{figure=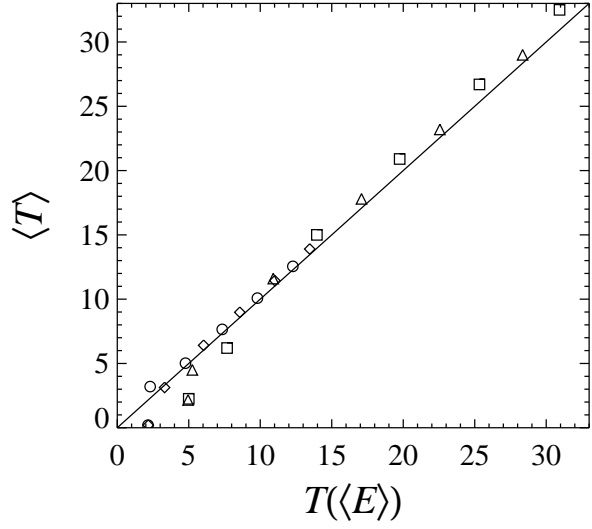,width=0.6\textwidth}
\end{center}
\caption{\label{fig7} Plot of the average period $\langle T \rangle$ plotted
versus the theoretical value $T(\langle E \rangle)$ computed using the value of
$\langle E \rangle$ obtained numerically. For each value of $\mu$ and $\delta$
(same symbols meaning than in Fig. \ref{fig5}) we use values of $\epsilon$
ranging from $\epsilon=10^{-2}$ to $\epsilon=10^{-7}$.}
\end{figure}

From a more quantitative point of view, and according to the previous
picture, we can compute the mean period $\langle T \rangle$, which in the purely
Hamiltonian case was a function of $E$, see Eq. (\ref{te2}), by using  the same function applied to
the mean value of $E$, i.e. $\langle T \rangle =T(\langle E  \rangle )$. This
relation has been checked in the numerical simulations. In Fig. \ref{fig7}
we plot the mean period $\langle T \rangle$ versus the period calculated from
the mean energy, $\langle E  \rangle$, which has also been evaluated
numerically. From this figure it appears that our qualitative argument of a
trajectory stabilized around the Hamiltonian orbit, corresponding to the
average energy, is well supported by the numerical simulations. 

In order to proceed further, we consider the probability distribution for the
amplitude variables, $P(A_1,A_2,A_3;t)$ which obeys a Fokker-Planck equation
\cite{FP}. For a general dynamics of the type given by Eq. (\ref{dotap}), it is
possible to show \cite{SMT} that the stationary probability distribution for
the $A_j$ variables is given by
\begin{eqnarray}
P_{\rm st}(A_1,A_2,A_3)& =& Z^{-1}\exp[-V(A_1,A_2,A_3)/\epsilon], \label{pdf} \\
Z & =&\int dA_1dA_1^\ast dA_2dA_2^\ast dA_3dA_3^\ast\, {\mathrm e}^{-V/\epsilon} 
\nonumber 
\end{eqnarray}

\noindent whenever two conditions are satisfied:
\begin{enumerate}
  \item[a)] Orthogonality condition (\ref{ortho}).
  \item[b)] The residual dynamics [nonrelaxational part of (\ref{dotap})] is
  divergence free:
  \begin{equation}\label{diver}
    \sum_{j=1}^3 \frac{\partial v_j}{\partial A_j}=0.
  \end{equation}
\end{enumerate}

\noindent In our case of the Busse-Heikes equations the orthogonality condition
is satisfied for $\mu=0$, $\delta>0$, and (\ref{diver}) is satisfied
independently of $\mu$ and $\delta$. For $\mu >0 $ this is no longer true but
we expect that for small $\mu$ a relation similar to (\ref{pdf}) would be valid
if we replace $V$ by a function $\Phi$ that differs from $V$ in terms that vanish for
vanishing
$\mu$.  Using this probability distribution, one can compute the average value
of the variable $E$ as:
\begin{equation}
\langle E \rangle = Z^{-1}\int dA_1dA_1^\ast dA_2dA_2^\ast dA_3dA_3^\ast\, E
\exp[-\Phi/\epsilon].
\end{equation}
We take the crude approximation $\Phi=V$ and, after a change of variables to
amplitude and phase, the mean value of the energy can then be computed as:
\begin{equation}
\langle E \rangle = \frac{\int_0^{\infty} dR_1 \int_0^{\infty}dR_2 
\int_0^{\infty}dR_3 E \exp[-V/\epsilon]}
{\int_0^{\infty}dR_1 \int_0^{\infty}dR_2 \int_0^{\infty}dR_3 
\exp[-V/\epsilon]},
\end{equation}
where $V$ and $E$ are given in terms of the variables $R_1,R_2,R_3$ in
Eqs. (\ref{potential}) and (\ref{defet}), respectively. In the case $\mu=0$ (for
which the above expression is exact) we obtain the value $\langle E \rangle =
1/60$, independent of $\epsilon$, and  $T=T(\langle E \rangle = 1/60) \approx
6.4467/\delta$. 

\begin{figure}[h]
\begin{center}
\psfig{figure=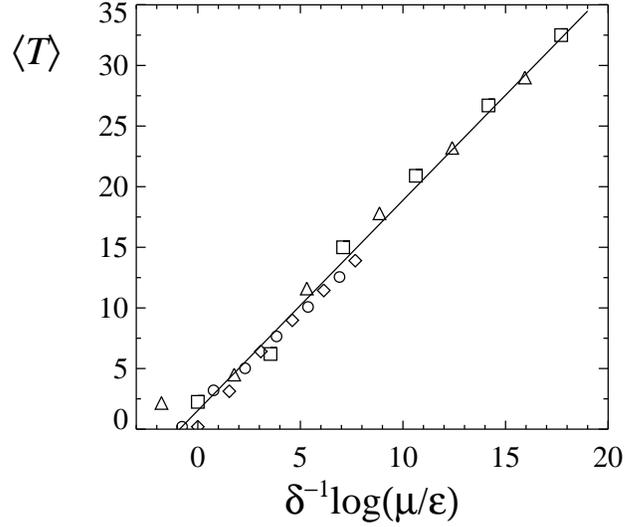,width=0.6\textwidth}
\end{center}
\caption{\label{fig8}  Average period, $\langle T \rangle$, plotted as a
function of $\delta^{-1}\log(\mu/\epsilon)$ in order to check the predicted
linear relation (\ref{tdme}). The straight line is the best fit and has a
slope of $1.73$. Same symbols meanings than in Fig. \ref{fig5} and values
of $\epsilon$ ranging from $\epsilon=10^{-2}$ to $\epsilon=10^{-7}$.}
\end{figure}

In the case $\mu>0$, the above integral can be performed by means of
a steepest descent calculation, valid in the limit $\epsilon \to 0$, where it yields the
asymptotic behavior $\langle E \rangle \to (\epsilon/\mu)^2$. The mean period can now be computed, in this limit of small $\epsilon$, using (\ref{te}), with the result that the period, as a function of the system parameters $\delta,~\mu,~\epsilon$, behaves as:
\begin{equation}
\label{tdme}
T(\epsilon,\mu,\delta) \approx \frac{3}{\delta}\ln(\mu/\epsilon),
\end{equation}

\noindent a relation that is expected to hold in the limit of small $\epsilon$
and for small values of $\mu$. The dependence with $\epsilon$ is
the same than the one holding for the mean first passage time in the decay from
an unstable state \cite{SMT} and also follows from the general arguments of 
\cite{stone}. In Fig. \ref{fig8} we show that there is indeed a
linear relation between the period computed in the numerical simulations and 
$\delta^{-1}\ln(\mu/\epsilon)$, as predicted by the above formula, although the
exact prefactor $3$ is not reproduced. We find it remarkable that, in view of
the simplifications involved in our treatment, this linear relation holds for a
large range of values for the parameters $\mu$, $\delta$ and $\epsilon$.

\section{Spatial-dependent terms}
\label{IV}

Tu and  Cross \cite{CrossTu} have proposed an alternative explanation for the
stabilization of the period without the necessity of the inclusion of the noise
terms: they modify the  Busse-Heikes equations by considering two-dimensional
amplitude fields, $A_j({\mathbf r},t)$, and including terms accounting for the
spatial variation of those fields:
\begin{eqnarray}
\partial_t A_1 & = & {\cal L}_1A_1+A_1 [1-|A_1|^2-(1+\mu+\delta)|A_2|^2-
(1+\mu-\delta)|A_3|^2],  \nonumber\\
\partial_t A_2 & = & {\cal L}_2A_2+A_2 [1-|A_2|^2-(1+\mu+\delta)|A_3|^2-
(1+\mu-\delta)|A_1|^2], \label{eq:spatial}\\
\partial_t A_3 & = & {\cal L}_3A_3+A_3 [1-|A_3|^2-(1+\mu+\delta)|A_1|^2-
(1+\mu-\delta)|A_2|^2].\nonumber 
\end{eqnarray}

\noindent Here ${\cal L}_j$ ($j=1,2,3$) are linear differential operators. Two
main classes of operators can be considered: isotropic and anisotropic. Whereas
a multiple scale analysis of the convective instability usually leads to
anisotropic terms, the isotropic terms are often justified for the sake of
mathematical and numerical simplicity. There are also the natural choice in
problems of population dynamics\cite{bennaim}. The simplest isotropic terms are the Laplacian
operators:
\begin{equation}
\label{ID}
{\cal L}_j^{\rm I}=\nabla^2, \quad j=1,2,3.
\end{equation}

Two types of anisotropic terms have been proposed for similar fluid problems in
the literature: (i) the Newell-Whitehead-Segel (NWS) terms \cite{NWS} and (ii)
the Gunaratne-Ouyang-Swinney (GOS) terms \cite{GOS,Graham}. Without altering
the essentials of the problem, both NWS and GOS terms can be further simplified
leading to second-order directional derivatives along three directions with a
relative orientation of $60^\circ$~\cite{CrossTu,echebarria}: 
\begin{equation} 
\label{AD} {\cal L}_j^{\rm A}=(\hat{\mathbf
e}_j\cdot\nabla)^2, \quad j=1,2,3, 
\end{equation} 
which are the only anisotropic terms considered henceforth. These are more
tractable numerically and will be used to compute the alternating period as
explained below.  

In this section we will compare the dynamical evolution corresponding to each
one of the isotropic and anisotropic spatial dependent terms presented before,
Eqs. (\ref{ID}) and (\ref{AD}), respectively.

Common to all of them is that,
as in section \ref{II}, we can recast system (\ref{eq:spatial}) into the form:
\begin{equation}\label{eq:BH_model2}
\partial_tA_j({\mathbf r},t)=-\frac{\delta{\cal F}_{\mathrm BH}}{\delta 
A_j^\ast}+\delta v_j,\quad j=1,2,3,
\end{equation}
where ${\cal F}_{\mathrm BH}$ is a real functional of the fields
given by:
\begin{eqnarray}\label{eq:BH_FBH}
  {\cal F}_{\mathrm BH}[A_1,A_2,A_3]& = &\int d{\mathbf r}\biggl[
  \sum_{j=1}^{3} \left(\frac{1}{2}|{\cal L}_j^{1/2}A_j|^2-
  |A_j|^2+\frac{1}{2}|A_j|^4 \right) 
  + \nonumber \\
  & &  (1+\mu)(|A_1|^2|A_2|^2+|A_2|^2|A_3|^2+|A_3|^2|A_1|^2\biggr]
\end{eqnarray}
and the functions $v_j$ are given by (\ref{vi}).

As in the zero-dimensional case of sections \ref{II} and \ref{BHnoise},
$\delta=0$ entails a relaxational gradient type dynamics and ${\cal F}_{\mathrm
BH}$ acts as a Lyapunov functional that decreases monotonically with time.
Since this potential is minimized by homogeneous solutions (because the
spatial-dependent term gives always a positive contribution) the stationary
solutions (and their stability) in the case $\delta=0$ are the same as in the
zero-dimensional case. Unfortunately, the orthogonality condition
\begin{equation}
  \delta\sum_{j=1}^3 \int d{\mathbf r}\,\frac{\delta{\cal F}_{\mathrm BH}}
  {\delta A_j}v_j+{\mathrm c.c.}=0
\end{equation}
is not trivially satisfied in the case $\mu=0$ for any of the spatial dependent
terms mentioned before, and the dynamical equations can not be reduced for
$\mu=0$ as in the zero-dimensional case. 

In general, for $0<\delta<\mu$, when the amplitudes grow from random initial
conditions around $A_j=0$, $j=1,2,3$, we expect the formation of interfaces
between the roll homogeneous states.  Those interfaces move due to curvature
and non-potential ($\delta > 0$) effects. Moreover, the fact of dealing with
three fields allows the formation of vertices, or points at which the three
amplitudes take the same value. In the potential case, $\delta=0$, the
interface motion is such that a final state in which a unique roll solution
fills the whole space is obtained (a process defined as ``coarsening"). On
the other hand, the nonpotential dynamics induces the rotation of front lines
around vertices giving rise to the formation of rotating spiral structures
\cite{GallegoCPC}. Similar structures have been observed in other three
competing species systems, such as lattice voter models \cite{porto}. For small
values of $\mu$, the interfaces are wide (it can be shown that an interface
varies over a length scale of order $1/\sqrt{\mu}$) and the density of vertices
is low. For large $\mu$ the interfaces are sharp and the density of vertices
increases. The exact shape of the spirals depends upon the spatial-derivative
terms used. With the isotropic terms, Eq. (\ref{ID}), interface propagation
follows the normal direction at each point so that closed domains have
spherical shape and spiral structures are close to Archimedes' spirals.  On the
other hand, for anisotropic spatial derivatives, Eq. (\ref{AD}), interface
propagation no longer follows the normal direction and closed domains stretch
or collapse along preferential directions so that they adopt an elliptic shape
rather than a spherical one. 

An important effect is that the rotation of interfaces around vertices, driven
by nonpotential effects, prevents the system from reaching a single roll
solution filling the whole space, even outside the K\"uppers-Lortz instability
region, i.e. for $\delta < \mu$ \cite{GallegoBH2d}. While this is true both for
isotropic and anisotropic derivatives, the dynamical mechanism that prevents
this coarsening is different for isotropic and anisotropic terms. For the
isotropic terms, vertices of opposite sense of rotation annihilate initially
with each other if located closer than a critical distance
$d_c\sim\delta^{-1}$. After a transient time in which vertices are formed, they
place each other outside the range of effective attraction of other vertices so
that their number is essentially constant, thus preventing coarsening. For the
anisotropic terms, two interfaces associated with the same vertex (and thus
rotating in the same sense) may collide and generate continuously new vertices
which, in turn, annihilate against each other
again preventing coarsening outside the instability region. A consequence of
interface motion is that a fixed point in space sees a change of the dominating
amplitude. This alternation change is essentially periodic in time and presents
a characteristic period which has nothing in common with the K\"uppers-Lortz
instability mechanism in the bulk. Therefore, the period associated to this
rotation is continuous at $\delta=\mu$, the instability point.

Before discussing what happens when this interface motion appears together with
the instability in the bulk, we mention that for the isotropic terms it is
possible to establish an analytical result concerning the front and spiral
motion. In this case, using the fact that interfaces move in the normal
direction to each point, it is possible to show  that the rotation angular
velocity of the interfaces around an isolated vertex scales, for small
$\delta$, as $\omega\sim\delta^2$ \cite{GallegoBH2d}. This predicts that, for
the isotropic derivatives, the average period in a fixed point of space coming
from the rotating spirals scales as $\langle T \rangle \sim \delta^{-2}$. 

As mentioned above, the mechanism of front motion due to the nonpotential
effects coexists with the K\"uppers-Lortz bulk instability. We will show in the
remaining of the section some results that follow, mainly, from a numerical
integration of Eqs. (\ref{eq:spatial}) in two spatial dimensions. It appears
from the numerical simulations that  the behavior beyond the instability point
(for $\delta>\mu$) depends strongly on the type of spatial derivatives used as
well on the magnitude of the parameter $\mu$. We discuss first each type of
derivatives separately.

\begin{figure}
\centering
\psfig{file=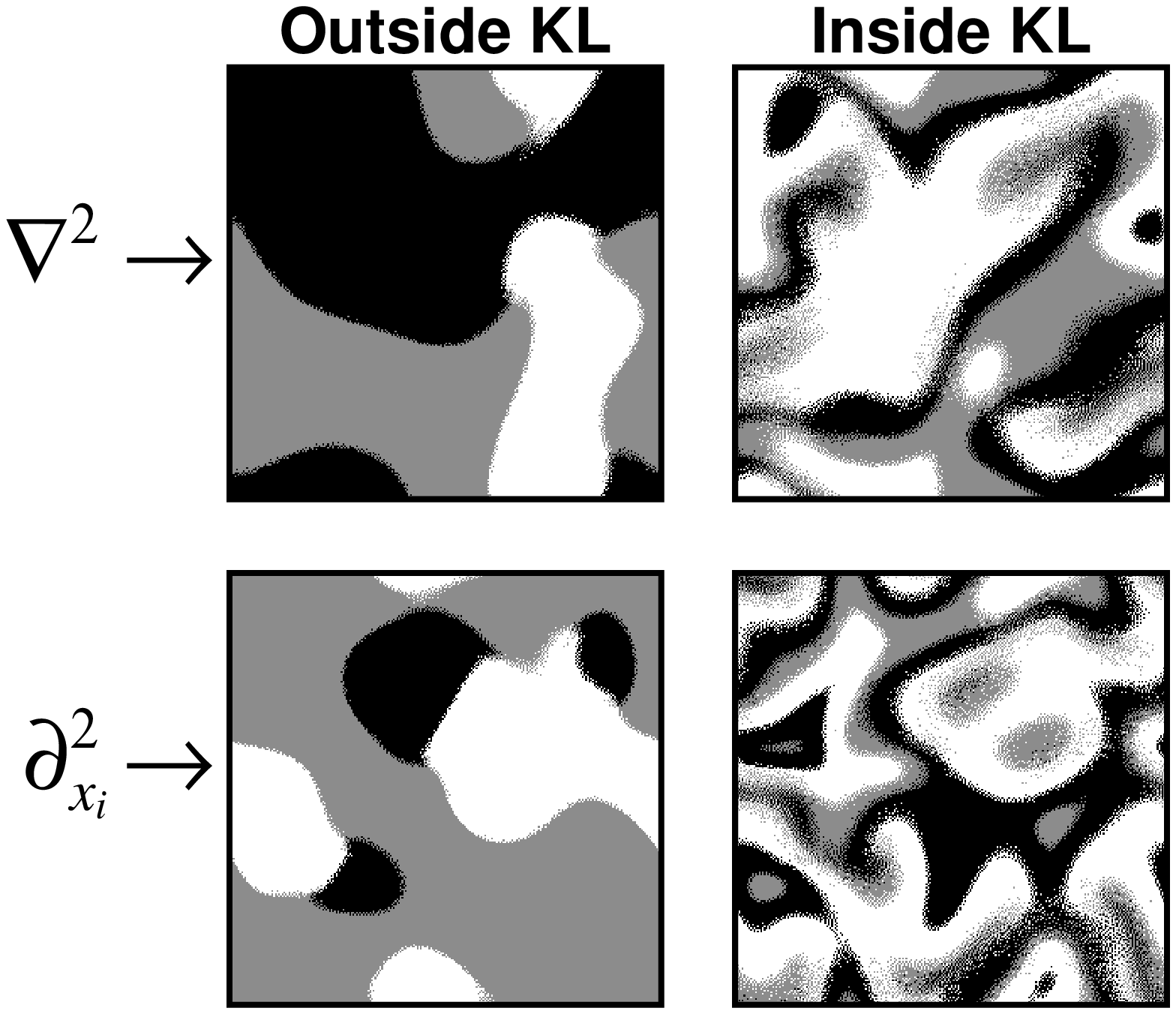,width=0.8\textwidth}
\caption{Four snapshots at long times corresponding to a numerical simulation
of the Busse-Heikes model [eq. (\ref{eq:spatial})] with isotropic (${\cal
L}^I_j=\nabla^2$) and anisotropic (${\cal L}^A_j=(\hat{\mathbf
e}_j\cdot\nabla)^2$) spatial
derivatives. Parameter values are: $\mu=0.1$ and $\delta=0.05$ ($1.3$) outside
(inside) the K\"uppers-Lortz instability region.
\label{fig:smallmu}}
\end{figure}

{\bf Isotropic derivatives}: For $\mu$ small, the bulk instability is such that
the intrinsic K\"uppers-Lortz period stabilizes to a statistically constant
value. In a given point of space, we can see that the dominant
amplitude changes due both to invasion from a rotating interface and a new
amplitude growing inside the bulk. We give evidence of this combined mechanism
in Fig. \ref{fig:smallmu} where we have used the value $\mu=0.1$ and we present
representative configurations inside and outside the instability region.

\begin{figure}
\centering
\psfig{file=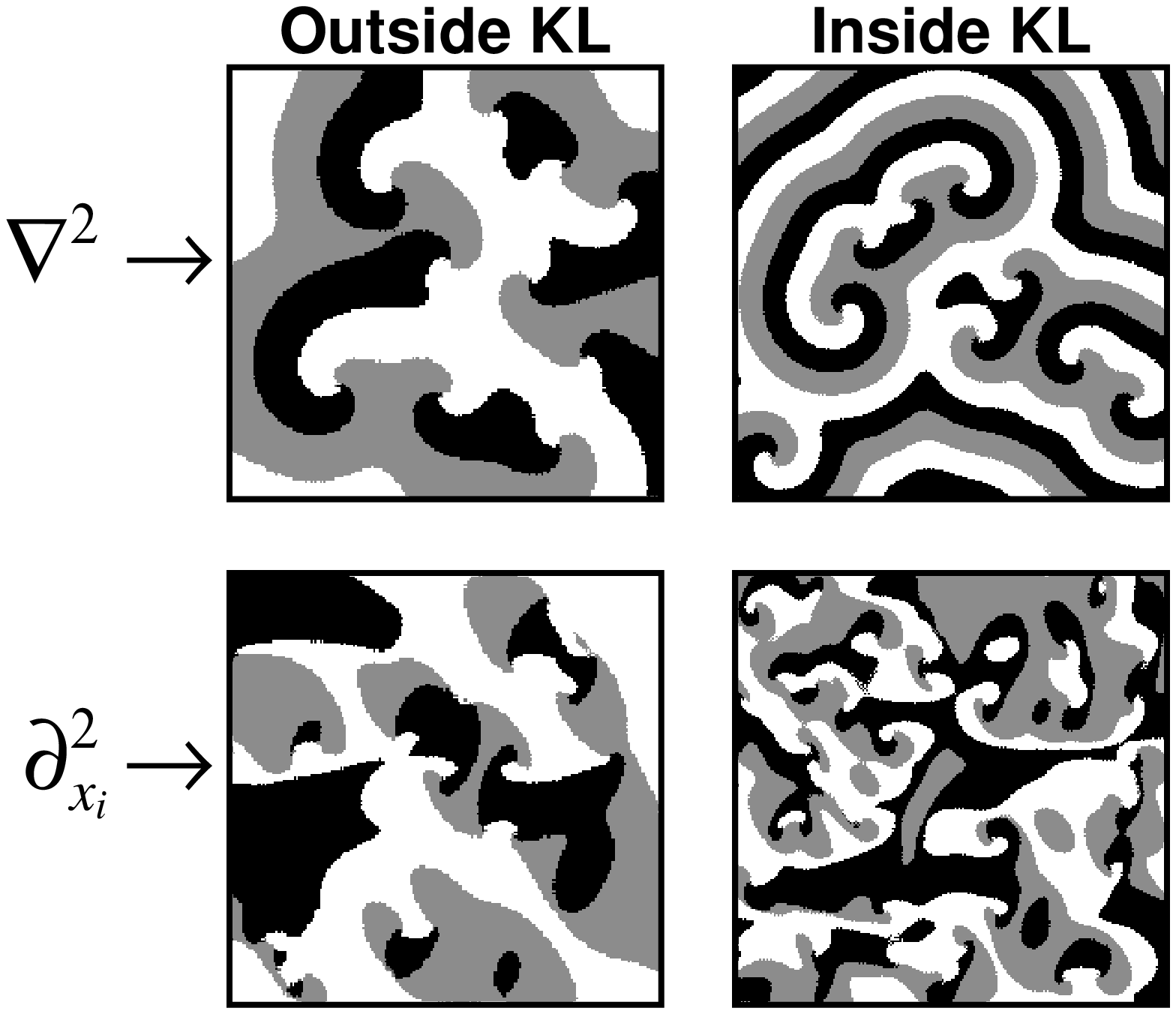,width=0.8\textwidth}
\caption{Four snapshots at long times corresponding to a numerical simulation
of the Busse-Heikes model [eq. (\ref{eq:spatial})] with isotropic (${\cal
L}^I_j=\nabla^2$) and anisotropic (${\cal L}^A_j=(\hat{\mathbf
e}_j\cdot\nabla)^2$) spatial
derivatives. Parameter values are: $\mu=2.5$ and $\delta=2$ ($3.5$) outside
(inside) the K\"uppers-Lortz instability region.
\label{fig:largemu}}
\end{figure}

For higher values of $\mu$, the K\"uppers-Lortz intrinsic period in the bulk is
observed to increase with time. This is the same phenomenon that occurs in the
zero-dimensional model without noise, see section II. Therefore at long times
the K\"uppers-Lortz period is so large that we only see rotating interfaces
around vertices, just like below the instability point. The two images of the
upper row in Fig. \ref{fig:largemu} show domain configurations at long times
for $\mu=2.5$, below ($\delta=2$) and beyond ($\delta=3.5$) the K\"uppers-Lortz
instability point in the case of the isotropic terms. Apart from the typical
size of the domains, it appears that there is no qualitative difference between
them.  The period of alternating amplitudes is entirely dominated by front
motion.

{\bf Anisotropic derivatives}: Both for small and large $\mu$, in the
K\"uppers-Lortz regime, $\delta>\mu$, we observe, in addition to the front
motion, domains of one phase emerging in the bulk of other domains; this is
seen at all times, indicating that, at variance with the isotropic derivative
case, the period associated with the K\"uppers-Lortz instability does not diverge
with time. Evidence is given in Fig. \ref{fig:smallmu} for $\mu=0.1$ and
Fig. \ref{fig:largemu} for $\mu=2.5$, both figures showing results inside
and outside the instability region.

\begin{figure}
\centering
\psfig{figure=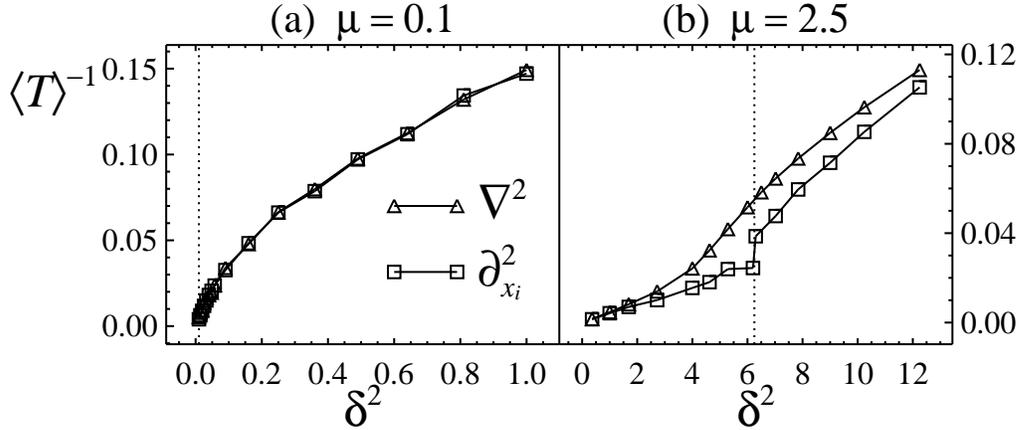,width=\textwidth}
\caption{Inverse of the alternating mean period as a function of $\delta^2$ for
the two-dimensional Busse-Heikes model with isotropic and anisotropic
spatial-dependent terms. We have chosen the coordinates in order to emphasize
the linear relation between the inverse of the period and $\delta^2$ valid for
small $\delta$\cite{GallegoBH2d}. Each plot
corresponds to a different value of the parameter $\mu$. The K\"uppers-Lortz
instability takes place at the right of the vertical dotted lines.
\label{fig:period}}
\end{figure}

For small $\mu$, in summary, the morphology of domains inside and outside the
instability region turns out to be similar with both kinds of spatial dependent
terms, Fig. \ref{fig:smallmu}. The alternating period for $\delta>\mu$ is
dominated by the K\"uppers-Lortz instability and is similar with isotropic and
anisotropic spatial derivatives. This shows up in the fact that the period
computed in a single point of space does not depend essentially of the type of
derivatives used, as shown in Fig. \ref{fig:period}a. 

For large $\mu$, on the
other hand, the morphology is different for isotropic and anisotropic terms.
For the isotropic ones, spiral rotation dominates the dynamics because of the
very large period associated with the bulk instability. For the anisotropic
terms, both front motion and bulk instability are present.
Finally in Fig. \ref{fig:period}b (large $\mu$) we show how the alternating
period changes when going through the K\"uppers-Lortz instability. We first note
that the period does not vanish in the stable regime ($\delta<\mu$). In this
regime it is entirely due to front and spiral motion. For isotropic derivatives
the period changes smoothly through the point $\delta=\mu$. This supports the
fact that the period is still given by front motion for $\delta>\mu$. On the
contrary, for anisotropic derivatives a jump in $T$ is observed at $\delta=\mu$.
In the K\"uppers-Lortz unstable regime and for anisotropic derivatives, $T$ is
determined by a combination of bulk instability and front motion.

\section{Conclusions}

We have analyzed the Busse-Heikes equations for Rayleigh-B\'enard convection in
a rotating fluid. For the situation of spatial-independent amplitudes, a case
previously analyzed by May and Leonard, we find a Lyapunov potential that
allows us, for $\mu=0$, to split the dynamics into a relaxational plus a residual
part. Since the residual dynamics is Hamiltonian, we are able to give explicit
relations for the time variation of the amplitudes and to compute the period of
the orbits as a function of the energy, which, in turn, is a function of
initial conditions. For $\mu>0$ we extend the previous picture by using an
adiabatic approximation in which the energy slowly decreases with time. This
allows us to compute the variation of the alternation period between the three
modes in the K\"uppers-Lortz instability regime.  We next consider the effect
of fluctuations and show how noise can stabilize the mean period to a finite
value. By using the Lyapunov potential employed in the deterministic case, we
can deduce an approximate expression that yields the period as a function of
the system parameters, $\mu,~\delta$ as well as a function of the noise
intensity $\epsilon$. The conclusion is that the period increases
logarithmically with decreasing noise intensity, a result that is well
confirmed by numerical simulations

The two-dimensional version of this problem exhibits rather different dynamical
behavior grossly dominated by vertices where three domain walls meet and which
have no parallel in lower dimensional systems. The rotation of interfaces
around vertices is driven by nonpotential effects and this inhibits coarsening
for sufficiently large systems. We investigated the influence on the dynamics
of the type of spatial dependent terms. For small values of the parameter
$\mu$, the morphology of domains inside the K\"uppers-Lortz region turns out to
be similar for both isotropic and anisotropic spatial derivatives. The
alternating period is dominated by the K\"uppers-Lortz instability and is
similar for both kinds of spatial-dependent terms. For large $\mu$, on the
contrary, the morphology of patterns as well as the alternating mean period are
different for isotropic and anisotropic terms. While the intrinsic period of
the instability diverges with time with isotropic derivatives, it saturates to
a finite value in the anisotropic case.

We acknowledge financial support from DGESIC (Spain) projects numbers
PB94-1167 and PB97-0141-C02-01.

\newpage

\end{document}